\documentclass{ckm}                 

\confname{Workshop on the CKM Unitarity Triangle, IPPP Durham, April
  2003}

\title{BTeV: STRATEGY AND SENSITIVITY}

\author{L.Moroni\thanks{Representing the BTeV collaboration.}}
\address{INFN, Milano}

\begin{document}

\begin{abstract}The next-generation experiments on Heavy Flavour Physics must
be performed at the hadron colliders to exploit the superior production cross
section and availability of all the species of $b$ and $c$ flavored hadrons. The
particularly hostile environment of the hadron colliders, on the other hand,
requires that new and adequate experimental tecniques be developed to face this
extraordinary challenge. The BTeV experiment has developed the right
experimental tools to perform this program with very high sensitivity and to 
extend our knowledge at the frontier of New Physics.
\end{abstract}

\maketitle


\section{Introduction}

In BTeV we believe that we have developed the right experimental techniques, 
both  instrumentally and for physical analysis, to deeply probe the $b$ and 
$c$ Heavy 
Flavour sector well beyond the possibilities of the present 
generation of experiments. Our goal is to look for evidence of  effects 
or inconsistencies brought in data by New Physics,  and thus be in a unique 
position to determine properties of the New Physics that can only be found 
by seeing how it modifies CP violation and rare decays.
 
Indeed, any discovery at LHC could impact the Heavy Flavour sector, 
specifically in modifying diagrams containing loops that contribute to 
CP violation and rare decays. Should this physics not be discernable, 
it would place great restrictions upon it, thus making precision measurements 
incredibly important. A large number of papers ($>$100), discuss specific 
examples. We also want to understand connections between quark mixing and 
neutrino mixing. One possible link \cite{masiero} within a SUSY GUT context 
predicts changes in specific CP violating angles. In the end, high precision 
measurements in the Heavy Flavour sector are and will be crucial and 
unavoidable for progress in our field, whatever the future scenario of 
particle physics turns out  to be. 

BTeV will perform this program in the forward region of the $p-\bar p$ 
collisions at the TeVatron collider at Fermilab. The experiment is 
designed to fully exploit the 
superior potential of hadron colliders in terms of number of  produced  
$b$-particles and, in particular, the advantages provided by working in 
the forward region at high rapidity. Here the high Lorentz boost, with 
which the $b-\bar b$ pairs are produced, drastically reduces multiple-scattering 
in the detector and the strong angular correlation between the two 
$b$-particles, dramatically enhances the flavour-tagging efficiency.

BTeV will be placed in the C0 experimental hall of the TeVatron collider 
and is expected to start its data taking run in 2008-2009. The BTeV program 
has been already approved by the Fermilab PAC, its base-line cost has been 
successfully reviewed by Fermilab and, now, the experiment is waiting for 
funding from the U.S. Department of Energy.

\section{Main experimental challenges}

To fully exploit the enormous advantages of doing future Heavy Flavour 
Physics in the forward region of the hadron colliders, adequate strategies 
must be adopted to face the following experimental challenges (typical of 
a high-luminosity hadron-collider environment).
\begin{enumerate} 
\item	One has to isolate the b-signal from a large total rate: 
$\sigma_{bb} /\sigma_{tot} \sim$ 1/500 at the Tevatron energy;
\item	Background from $b$-decays can overwhelm "rare" processes; 
\item	Large data rates are expected just from $b$-decays: ~1 KHz within the 
acceptance;
\item	Large particle rates cause radiation damage to the detectors and 
high photon multiplicities may obscure the signal in the EM calorimeter.
\end{enumerate}

\begin{figure*}
\hbox to\hsize{\hss
\includegraphics[width=.5\hsize]{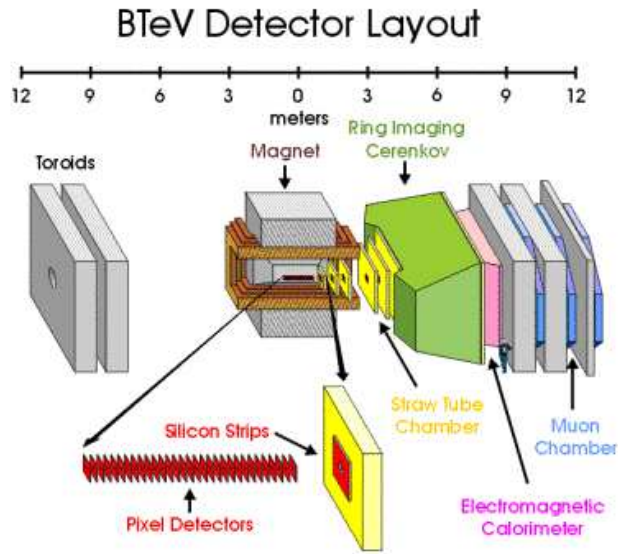}
\hss}
\caption{The BTeV detector.}
\label{fig:spectrometer}
\end{figure*}

The BTeV approach \cite{prop,update} is characterized by the following key elements, 
which specifically address and resolve the previous experimental 
challenges.
\begin{enumerate}
\item	High Precision silicon pixel vertex detector to provide excellent 
reconstruction of primary (interaction), secondary, and tertiary vertices
\cite{pixel};
\item	First Level Trigger on detached vertices to select $b$-signals and 
reject background;
\item	Excellent Particle \& Lepton ID;
\item	Dead-timeless trigger and DAQ capable of writing KHz of events 
to tape;
\item	PbWO$_4$ crystal calorimeter \cite{EM}.
\end{enumerate}
This approach is implemented through a suitable choice of detectors as 
illustrated in Fig.~\ref{fig:spectrometer}.
 
 The experimental apparatus is designed around a pixel vertex detector 
 which is placed inside the central dipole magnet centered on the interaction 
 region. The pixel detector provides fast and precise hit information to the 
 trigger processors that make a trigger decision on the evidence for decay 
 vertices in the event. The forward tracking system consists of seven 
 stations of straw tube chambers (in yellow) with silicon micro-strip 
 detectors (in red) in the innermost region, close to the beam pipe. 
 Each station provides three coordinate measurements, X, along the 
 horizontal direction, U and V, at  $\pm$11$^o$ around the vertical direction. 
 The forward tracking system improves the momentum resolution of the tracks 
 reconstructed by the vertex detector, allows for the reconstruction of 
 the neutral Vees decaying downstream of the vertex detectors and measures 
 with precision the trajectories of the tracks crossing the RICH detector 
 (in green). At the end of the spectrometer are the EM calorimeter (in pink) 
 and the muon detector, consisting of an array of three iron filters (in grey) 
 and three stations of muon chambers (in blue). The first two iron blocks 
 are magnetized with a toroidal field.

The apparatus is very compact and can be easily upgraded to cover the 
opposite hemisphere of rapidity. In the next section of this paper I 
will concentrate on the key element of the BTeV apparatus, the vertex trigger.

\section{The BTeV vertex trigger}

The BTeV vertex trigger can be viewed as the heart of the experiment.  
Every single crossing, at 7.6 MHz frequency, is analyzed  to search for 
evidence of detached vertices. 
This is done by reconstructing the tracks from the collision, finding 
the  primary interaction vertex,  and then determining the number of 
tracks missing the primary vertex and having an impact parameter with 
a significance above a certain threshold value: $b/\sigma_b > n$ , where $b$ 
is 
the measured impact parameter and $\sigma_b$ its corresponding error. 
All this is made possible by the unique features provided by the pixel 
vertex detector fully immersed in a magnetic field. The very low occupancy, 
together with the excellent spatial resolution and the intrinsically 3D 
coordinate information, makes the track reconstruction process very simple 
and practically free of fake track combinatorics. Thanks to the presence 
of the magnetic field, the track momenta can be simultaneously measured 
and the low momentum tracks disregarded. These tracks have large multiple 
scattering, which can cause false impact parameters leading to a poor 
background rejection in the trigger. Using our pixel prototype detectors 
in a test beam at Fermilab, we measured resolutions ranging from 
4 to 10 $\mu$m, depending on the track crossing angle, 
and expect a fake tracks at the trigger level with a 
probability much lower than 1\%. 
The proximity of the pixel planes to the beam line, 6 mm 
in our case, together with the pixel resolution, 
determines the impact parameters resolution and hence drives the trigger 
performance.
The system of processors, used to run the trigger software, is based on 
a heavily pipelined and parallel architecture using inexpensive processing 
nodes optimized for specific tasks and sufficient memory to buffer the 
event data while calculations are carried out. The system employs about 
3000 processors, DSPs \& FPGAs, and $\sim$1 Terabyte buffer memory.

There are several crucial advantages of using the detached vertex trigger at 
the first level. First of all, it allows us to select the most general class 
of $b$ events without any bias toward particular final states. In particular, 
since it performs essentially the same selection that will be employed in 
the off-line analysis to isolate the signals, its effective efficiency, 
expressed as (number of reconstructed events) /( number of reconstructable 
events), turns out to be very high. It is worth noting that this ratio 
represents the figure of merit that counts in the end. Furthermore, the 
same trigger allows for an extensive charm physics program at very high 
sensitivity. 
The trigger algorithm, being completely programmable, can be configured 
in a variety of different ways. Unexpected situations, which could arise 
from the unknown characteristics of the extreme forward regions of the 
hadron colliders, can be easily handled thanks to this feature.  
The base-line algorithm has been developed to present an excellent 
robustness even in presence of more interactions per bunch crossing. 
At the BTeV nominal luminosity, $L$ = 2 $\times$ 10$^{32}$ cm$^
{-2}$s$^{-1}$, two interactions 
on average are produced per bunch crossing, but the algorithm performance 
has been proven to remains unchanged by no more than 20\%, up to 6 
interactions per bunch crossing.  Table~\ref{table:trigeff} shows the 
simulated trigger 
efficiency with our base-line algorithm for a wide class of $b$-decays. 
The algorithm performance has been tuned to 
reject 99\% of the Minimum Bias events.

\begin{table}[h!]
\begin{center}
\begin{tabular}{lc} 

Process & Eff.(\%)\\ 
\hline\hline
Minimum Bias&1\\
$B_s \rightarrow D_sK^-$&74\\
$B^o \rightarrow D^{*+}\rho^-$&64\\
$B^o \rightarrow \rho^o \pi^o$&56\\
$B^o \rightarrow J/\psi K_s$&50\\
$B^o \rightarrow J/\psi K^{*o}$&68\\
$B^- \rightarrow D^oK^-$&70\\
$B^- \rightarrow K_s \pi^-$&27\\
$B^o \rightarrow 2-body$&63\\
$(\pi^+\pi^-,K^+\pi^-,K^+K^-)$&\\
\end{tabular}
\caption{BTeV trigger efficiencies for a wide class of $b$-decays.}
\label{table:trigeff}
\end{center}
\end{table}

\section{Physics reach of BTeV}

The sensitivity of BTeV to the CP violating angles is summarized in 
Table~\ref{table:ckmsensi} for a luminosity of  2 $\times$ 10$^{32}$ cm$^
{-2}$s$^{-1}$ and 
10$^7$ seconds of running 
time \cite{BTeV}. The performance is outstanding and covers all the 
spectrum of the 
important measurements. It worth noting that the sensitivity to the 
$\alpha$-angle through $B^o \rightarrow \rho \pi$, as well as that 
to $\chi$ through $B_s \rightarrow J/\psi \eta^{(')}$, is very high.

\begin{table*}[tb]
\begin{center}
\begin{tabular}{llllll} 

Reaction & BR(10$^{-6}$)&\# of Events & S/B & Parameter &Error or (Value) \\ 
\hline\hline
$B^o \rightarrow \pi\pi$&4.5&	14,600&	3&	Asymmetry&	0.030 \\ 
\hline
$B_s \rightarrow D_sK^-$&300&	7,500&	7&$\gamma$ - 2$\chi$&	8$^o$ \\ 
\hline
$B^o \rightarrow J/\psi K_s$&445&	168,000	&10&	sin(2$\beta$)&	0.017\\ 
\hline
$B_s \rightarrow D_s\pi^-$& 3000	&59,000	&3	&$x_s$	&(75)\\
\hline
$B^- \rightarrow D^o(K^+\pi^-)K^-$&  0.17	& 170	& 1&  & \\
$B^- \rightarrow D^o(K^+K^-)K^-$& 1.1	&1,000	&$>$10	&$\gamma$	&13$^o$  \\
\hline
$B^- \rightarrow K_s\pi^-$&  12.1	& 4,600	& 1	&& theory errors  \\
$B^o \rightarrow K^+\pi^-$& 18.8	&62,100	&20&$\gamma$&$<$ 4$^o$  \\
\hline
$B^o \rightarrow \rho^+ \pi^-$&28	&5,400	&4.1 & & \\
$B^o \rightarrow \rho^o \pi^o$&5	&780	&0.3	&$\alpha$	&~4$^o$   \\
\hline
$B_s \rightarrow J/\psi \eta$& 330	&2,800	&15 & &\\
$B_s \rightarrow J/\psi \eta^{\prime}$ &670	&9,800	&30	&sin(2$\chi$)&	0.024  \\

\end{tabular}
\caption{BTeV physics reach on CKM parameters in 10$^7$ seconds. Reactions 
between the same horizontal lines are simultaneously used to measure the same 
parameter.}
\label{table:ckmsensi}
\end{center}
\end{table*}

BTeV has an excellent reach in rare decays too, as shown in Table~
\ref{table:raresensi} 
together with the physics issue that each measurement addresses.

\begin{table*}[tb]
\begin{center}
\begin{tabular}{lllll} 

Reaction & BR(10$^{-6}$)&\# of Events & S/B & Physics \\ 
\hline\hline
$B^o \rightarrow K^{*o}\mu^+\mu^-$& 1.5	&2,530	&11&polarization $\&$ rate\\ 
\hline
$B^- \rightarrow K^- \mu^+\mu^-$&0.4&	1,470&	3.2&	rate\\ 
\hline
$B \rightarrow s \mu^+\mu^-$& 5.7&4,140	&0.13&rate: Wilson coefficents \\

\end{tabular}
\caption{BTeV physics reach on rare decays in 10$^7$ seconds.}
\label{table:raresensi}
\end{center}
\end{table*}

Precision studies of b decays can bring a wealth of information to bear on 
new physics, that probably will be crucial in sorting out anything seen  
at LHC. The BTeV data sample will be large enough to test different 
scenarios emerging from "New Physics" at the TeV energy scale. 
Table~\ref{table:newpsensi} shows the expected rates in BTeV  for one year of running and 
in an $e^+e^-$ B-factory operating at the $\Upsilon$(4S) with a total 
accumulated 
sample of 500 fb$^{-1}$, about what is expected before BTeV begins running.

\begin{table*}[tb]
\begin{center}
\begin{tabular}{l|ccc|ccc} 

Mode & BTeV &(10$^7$ s)&&B-fact&(500 fb$^{-1}$)& \\ 
\hline\hline
 &Yield&Tagged&S/B&Yield&Tagged&S/B \\
\hline 
$B_s \rightarrow J/\psi \eta^{(\prime)}$&12650&1645 &$>$15&0& &	 \\ 
$B^- \rightarrow \phi K^-$&11000&11000&	$>$10&700&700& 4\\ 
$B^o \rightarrow \phi K_s$&2000	&    200&5.2&250&75& 4\\
$B^o \rightarrow K^{*o}\mu^+\mu^-$&2530	&  2530	&  11&~50& ~50&	 3\\ 
$B_s \rightarrow \mu^+\mu^-$&6&	     0.7&$>$15&	      0	&&\\
$B^o \rightarrow \mu^+\mu^-$&1&	     0.1&$>$10&	      0&&\\
$D^{*+}\rightarrow \pi^+ D^o(K^-\pi^+)$&$~$10$^8$&$~$10$^8$&large
&8$\times$10$^5$&8$\times$10$^5$&large\\

\end{tabular}
\caption{Sensitivities on new physics modes. Comparison of BTeV and 
B-factory yields on different time scales.}
\label{table:newpsensi}
\end{center}
\end{table*}

\section{Conclusions}

Heavy Flavour Physics is not only the ideal laboratory to investigate 
CP-violation and  probe the Standard Model at an unprecedented level, 
but it is even more crucial in revealing which kind of New Physics is 
hidden in the data. 
The present generation experiments, BaBar \& BELLE together with CDF, will 
produce a lot of very exciting results in the next few years.  
Starting in about 2008-2009, the high statistics results from BTeV and 
LHCb will take us to the next level, expanding our searches to a variety 
of other modes and fully engaging a major new player, Bs.

\end{document}